\documentclass[conference]{IEEEtran}
\IEEEoverridecommandlockouts

\usepackage{cite}
\usepackage{amsmath,amssymb,amsfonts}
\usepackage{algorithm}
\usepackage{algorithmic}
\usepackage{array}
\usepackage{booktabs}
\usepackage{graphicx}
\usepackage{subfigure}
\usepackage{textcomp}
\usepackage{xcolor}
\usepackage{graphicx}
\usepackage{tablefootnote}
\usepackage{multirow}
\usepackage{marvosym}
\usepackage{arydshln}
\usepackage{amsmath,graphicx,amsfonts,svg,float}
\usepackage{booktabs,multirow,stfloats}
\usepackage[colorlinks,
            linkcolor=blue,
            anchorcolor=blue,
            citecolor=blue]{hyperref}

\def\BibTeX{{\rm B\kern-.05em{\sc i\kern-.025em b}\kern-.08em
    T\kern-.1667em\lower.7ex\hbox{E}\kern-.125emX}}
\begin{document}


\title{\textsc{\textsc{ConTuner}}: Singing Voice Beautifying \\with Pitch and Expressiveness Condition}


\author{
\IEEEauthorblockN{
Jianzong Wang$^{1}$, 
Pengcheng Li$^{1,2}$ , 
Xulong Zhang$^{1\textsuperscript{\Letter}}$ \thanks{$\textsuperscript{\Letter}$ Corresponding author.},  
Ning Cheng$^{1}$, 
Jing Xiao$^{1}$}
\IEEEauthorblockA{
\textit{$^{1}$Ping An Technology (Shenzhen) Co., Ltd.} \\
\textit{$^{2}$University of Science and Technology of China}}
\IEEEauthorblockA{
\texttt{jzwang@188.com, lipengcheng@ustc.edu,
zhangxulong@ieee.org, }
\\ \texttt{\{chengning211, xiaojing661\}@pingan.com.cn
}}}

\maketitle

\begin{abstract}
 Singing voice beautifying is a novel task that has application value in people's daily life, aiming to correct the pitch of the singing voice and improve the expressiveness without changing the original timbre and content. Existing methods rely on paired data or only concentrate on the correction of pitch. However, professional songs and amateur songs from the same person are hard to obtain, and singing voice beautifying doesn't only contain pitch correction but other aspects like emotion and rhythm. Since we propose a fast and high-fidelity singing voice beautifying system called \textbf{\textsc{ConTuner}}, a diffusion model combined with the modified condition to generate the beautified Mel-spectrogram, where the modified condition is composed of optimized pitch and expressiveness. For pitch correction, we establish a mapping relationship from MIDI, spectrum envelope to pitch. To make amateur singing more expressive, we propose the expressiveness enhancer in the latent space to convert amateur vocal tone to professional. \textsc{ConTuner} achieves a satisfactory beautification effect on both Mandarin and English songs. Ablation study demonstrates that the expressiveness enhancer and generator-based accelerate method in \textsc{ConTuner} are effective.
\end{abstract}


\section{Introduction}
Singing voice beautifying (SVB) is a novel task in the field of speech and music. SVB models aim to calibrate the pitch of the amateur singing voice while retaining the timbre and content of the original singing voice and then improve the singing skills and expressiveness of the singing voice. In the entertainment industry, SVB is often completed by vocal tuners using professional tools like Auto-Tune \cite{yong2018singing}, which requires expensive labour costs. Since many people enjoy singing but face difficulties to obtain satisfactory songs, automatic singing beautification has great application value in our daily lives.

There are currently two types of tasks associated with SVB, singing voice conversion (SVC) and automatic pitch correction (APC). SVC changes the singer of the source singing voice and maintains the content. But in SVB, we need to keep the content and the vocal timbre. APC directly corrects the pitch of amateur singing voices, but this is insufficient to perform beautification. Current SVB work \cite{nsvb} relies on paired data, but it is difficult to obtain amateur and professional songs from the same person. The data we can easily obtain are amateur singing voices and original professional recordings of the same song. Lots of improvements in previous works are made only on pitch correction, but few researchers focus on the expressiveness of a song like singing skill, emotion, rhythm, \textit{etc}. How to generate a new song with a newly generated pitch and beautified features is also a critical problem. One way is to use a vocoder such as WORLD \cite{2016world}, but this kind of generation work is difficult to establish the control of expressiveness and other features of the song. 

To address the mentioned problems, we propose a diffusion-based SVB model,  \textbf{\textsc{ConTuner}}, which integrates the predicted pitch information as well as the modified expressiveness into the original amateur singing voice to achieve singing voice beautification. As far as we know, we are the first to model the expressiveness in singing voices in the SVB area, and we hope to bring inspiration to future works. Furthermore, as an outstanding SVB model is expected to be fast, we choose the high-performance generative model \textit{Diffusion} \cite{df1} as the backbone of \textsc{ConTuner}. And inspired by some recent works, we speed up the diffusion model through using \textit{generator-based} methods \cite{diffgan,DBLP:conf/iclr/SalimansH22}. During the whole training and reasoning process of the model, \textsc{ConTuner} only needs to extract target information from amateur singing voice constraints, thereby avoiding the problem of paired data. In addition, \textsc{ConTuner} beautifies the amateur singing voice from the perspective of pitch and expressiveness. Moreover, \textsc{ConTuner} chooses to establish control of the condition during the generation of the beautified Mel-spectrogram instead of establishing control of the condition in the vocoder stage. To sum up, our contributions can be summarized as follows:

\begin{itemize}
    \item We propose a novel model called \textsc{ConTuner} to solve the task of beautification \textit{without} professional-amateur paired data from the same singer. We provide a new perspective that obtains conditions from pitch and expressiveness and establishes the control of the condition in the process of Mel-spectrogram generation. 
    \item We \textit{predict} pitch instead of fitting pitch in terms of pitch correction. We establish a mapping relationship from MIDI, spectral envelope to pitch curve to get the corrected pitch curve. We derive an expressiveness representation from Mel-spectrogram and obtain the beautified expressiveness via the designed expressiveness enhancer.
\end{itemize}

\begin{figure*}[!ht]
    \centering
    \includegraphics[width=0.86\textwidth]{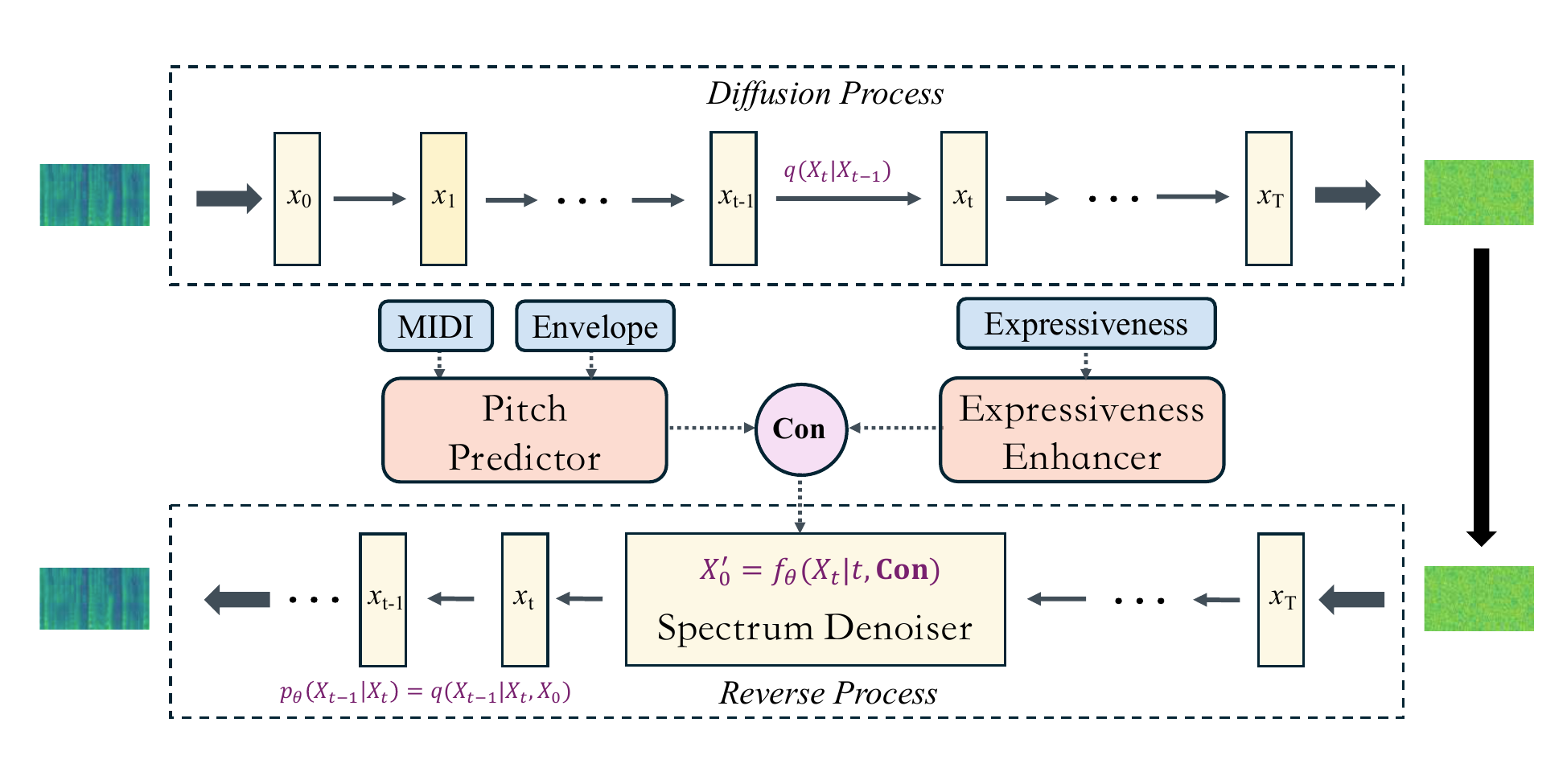}
    \caption{Architecture of \textsc{ConTuner}. The pitch predictor conducts mapping from MIDI and envelope to pitch, while the expressiveness enhancer disentangles the expressiveness representation from the singing voice. The outputs from them are combined as the condition that takes part in the denoising process.}
    \label{figure1}
\end{figure*}

\section{Related Works}
\subsection{Neural Singing Voice Processing}
Utilizing neural networks for singing voice processing is an area that has begun to attract attention in recent years \cite{sun2022Investigation,zhang2022SUSing}. Different from speech voice processing (like speech voice conversion \cite{jc2019one,autovc}, text-to-speech \cite{jonathan2018natural,yi2021fastspeech2}, \textit{etc}), singing voice contains richer and more varied pitch and rhythmic information, which requires more accurate control and optimization of these information. Existing works on neural singing voice processing contain automatic pitch correction \cite{zhou2009canonical,luo2018singing,zhuang2021karatuner}, singing voice conversion \cite{vc4,vc2}, and singing voice beautifying \cite{nsvb}. APC attempts to modify the unsatisfactory pitch of singing voices from amateur singers but APC doesn't further consider the beautification of the processed singing voice, such as singing skills, emotion, expressiveness, \textit{etc}. SVC is a downstream task of voice conversion, which aims to change the singer of a given singing voice. SVC methods usually first disentangle the timbre and the content from the target song and source song respectively, and then combine the timbre information from the target singer with the content information from the source song \cite{liu2021diffsvc,liu2021fastsvc,lu2020vaw}. SVC is quite different from SVB, as SVC will erase all the uniqueness of the singing voice while SVB tries to keep the content and the vocal timbre unchanged. Simply leveraging the SVC method for the SVB task will greatly destroy the original singers' unique characteristics. 

\subsection{Diffusion Model}
Denoising diffusion probabilistic model (DDPM) \cite{df1,nichol2021improved} gradually transforms simple distributions into complex data distributions with Markov chains. It has gained satisfactory performance in diverse tasks like image generation \cite{aditys2022hierarchical}, speech synthesis \cite{kong2021diffwave}, and even object detection \cite{chen2023diffusiondet}. The forward diffusion process and reverse generation process are built in DDPM to learn the transformation. In the forward process, Gaussian noise is gradually incorporated into the data, this helps the model to explore the various data distributions, while the model tries to denoise to restore the origin data through the reverse process. The reverse process makes the model gain the capability to generate real samples. Compared to the classic generated model, generative adversarial network (GAN), DDPM is more stable but relatively slow. How to improve the inference speed of DDPM also becomes an issue worth exploring \cite{lu2022dpm,ma2023deepcache,lu2022dpmp}.

\section{Methodology}
\subsection{Generator- and Condition-based Diffusion}
The overall architecture of \textsc{ConTuner} is shown in Fig. \ref{figure1}. Diffusion model is adapted as the backbone of \textsc{ConTuner}, combining with modified condition (\textit{i.e.} the output of the pitch predictor and expressiveness enhancer). Through the diffusion process and reserve process, we can get the Mel-spectrogram of the beautified singing voice. The diffusion process is non-parametric. We input the Mel-spectrogram $x_{0}$ of amateur song, and after $\mathit{t}$ steps of sampling, we get the Mel-spectrogram $x_{t}$ with noise. With the pre-defined noise schedule $\beta$ and diffusion step $\mathit{t}$, we compute the corresponding constants in Eq. \ref{eq:constant} respectively.
\begin{equation}
    \alpha _{t} = \prod_{i=1}^{t} \sqrt{1 - \beta _{i} }  \qquad  \sigma _{t}=\sqrt{1-\alpha _{t}^{2} }
\label{eq:constant}
\end{equation}

In conventional gradient-based training, the noise $\epsilon $ loss is calculated through optimizing a random term of $\mathit{t} $ with stochastic gradient descent as shown in Eq. \ref{eq:noiseloss}.
\begin{equation}
    \mathcal L_{\theta }^{G} =  \parallel \epsilon _{\theta } (\alpha _{t} x_{0} +\sqrt{1-a_{t}^{2}  }\epsilon  )-\epsilon \parallel_{2}^{2},\epsilon\sim \mathcal N(0,1)
\label{eq:noiseloss}
\end{equation}

It is well known that in DDPMs, $x_{t} $ has different degrees of perturbation, thus using a single gradient-based parameterization network directly in different $t$ predictions for $x_{t-1} $ is difficult. Inspired by the recent works \cite{diffgan,BDDM,huang2022prodiff} in the field of text-to-speech, which point out that the generator-based diffusion model does not need to estimate the gradient of the data density. So we predict clean professional Mel-spectrogram $x_{0}^{p}$ by undisturbed $x_{0}'$, and then add the anti-perturbation through the posterior distribution $q(x_{t-1}|x_{t} ,x_{0}')$

Therefore, in the reverse sampling process of the diffusion model, $x_{t-1}$ is sampled with the posterior distribution given $x_{t}$ and the predicted $x_{0}'$. 
\begin{equation}
    x_{t-1}\sim p_{\theta }(x_{t-1}|x_{t})=q(x_{t-1}|x_{t},x_{0}')
\end{equation}

In this way, \textsc{ConTuner} can greatly reduce the quantity of sampling steps $\mathit{t}$. We put the beautifying condition $Con$ into the denoiser to guide the direction of generation as shown in Eq. \ref{eq:con}.
\begin{equation}
    x_{0}'=f_{\theta }(x_{t}|t,Con)
\label{eq:con}
\end{equation}

Finally, we constrain the loss of the denoiser which is defined as a mean squared error (MSE) in the data space $\mathit{x}$. Efficient training is to optimize a random term of $t$ with stochastic gradient descent.
\begin{equation}
    \mathcal L_{\theta }^{Denoiser} =  \parallel x_{\theta } (\alpha _{t} x_{0} +\sqrt{1-a_{t}^{2}  }\epsilon  )-x_{0}^{p} \parallel_{2}^{2}  
\label{eq:denoiser}
\end{equation}

\subsection{Pitch Predictor and Expressiveness Enhancer}
\begin{figure}[t]
    \centering
    \includegraphics[width=0.48\textwidth]{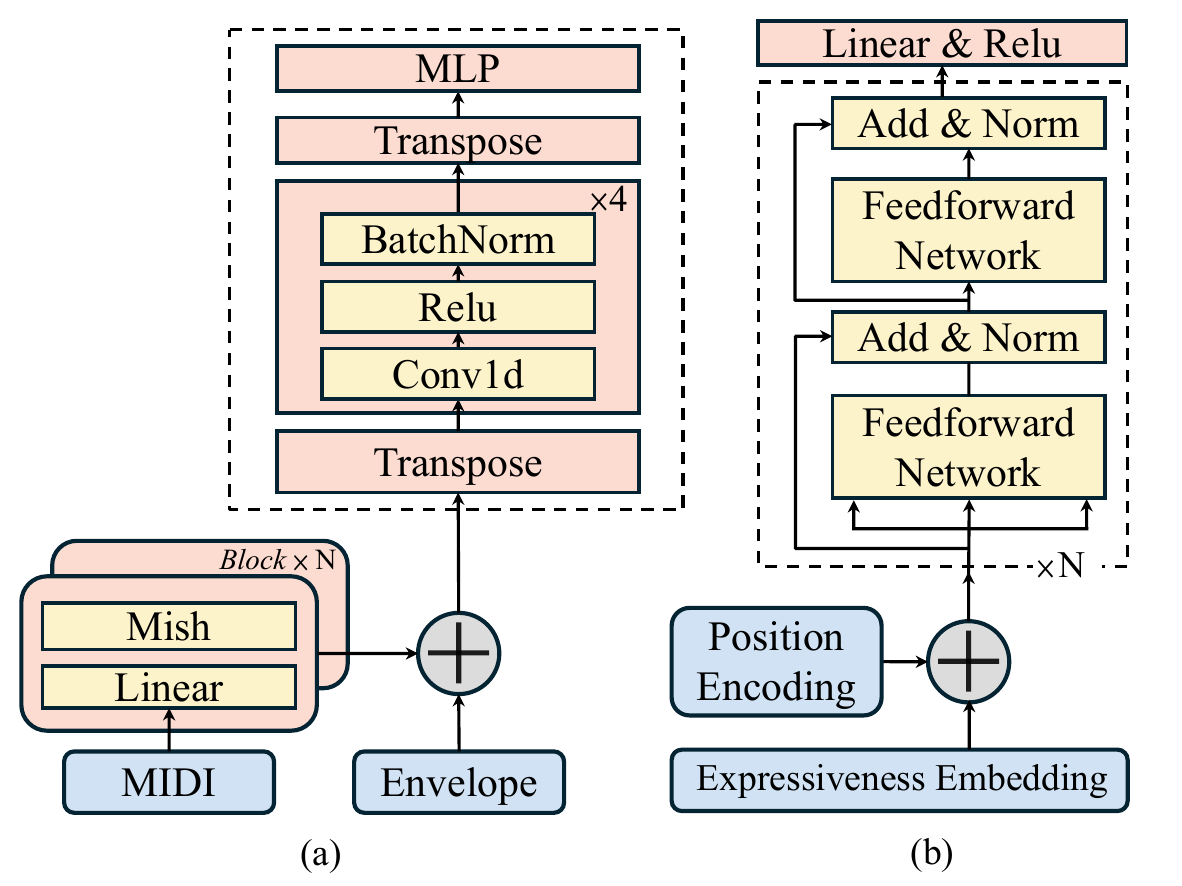}
    \caption{Details of the pitch predictor and expressiveness enhancer.}
    \label{fig:ppee}
\end{figure}

\subsubsection{Pitch Predictor}
The structure of the \textit{pitch predictor} (PP) is shown in Fig. \ref{fig:ppee}(a). As we hope to build the beautified pitch curve and maintain the pitch characteristics of the amateur singer at the same time, so we combine the spectral envelope and MIDI from professional singers and amateur singers respectively. We establish a mapping $pitch\leftarrow(envelope, MIDI)$. The spectral envelope feature implicitly contains the pitch curve \cite{2003new}, as previous work predicts the pitch curve from the spectral envelope with high accuracy. We feed spectral envelope into the designed PP. Besides, MIDI can be seen as the standard pitch representation of a song. We perform HMM smoothing to extract standard MIDI note sequences from vocals via \textit{pyin} \cite{midipyin}. With the help of the WORLD vocoder \cite{2016world}, we extract the spectral envelope and pitch curve (set as the pitch label $p$). The MIDI vector goes through 4 linear layers, each layer is followed by a mish activation function, then spliced and flipped with the spectral envelope dimension. The main body of PP contains four Conv1d layers with ReLU activation function and batch normalization followed, finally the predicted pitch curve $p'$ can be obtained from the MLP layer. 

During the training process of the model, we input the MIDI and the spectral envelope of amateur singers, while the labels are also extracted from amateur singing voices. It is worth noting that PP does not require any information of professional songs during the training stage. In the inference stage, we input the spectral envelope from professional singers and the MIDI from amateur singers to obtain the beautified pitch curves.

\subsubsection{Expressiveness Enhancer}
The structure of \textit{expressiveness enhancer} (EE) is shown in Fig. \ref{fig:ppee}(b). We define the singing skills, rhythm, and emotion of a singing voice as the expressiveness characteristics. We first obtain expressiveness representation through disentanglement similar to voice conversion. Defining a pair of expressiveness representations $(W_{a},W_{p})$, which represent the amateur expressiveness $W_{a}$ and professional expressiveness $W_{p}$ of the same song respectively. We disentangle expressiveness representation with a pre-trained encoder. A feed-forward Transformer block is adapted as the main body of the expressiveness enhancer, which is a stack of self-attention layers \cite{attention}. The latent expressiveness representation $W_{a}'$ represents the expressiveness after being modified by EE. What we need to do is to strengthen the similarity between $W_{a}'$ and $W_{p}$.

\subsubsection{Condition Loss}
Condition loss consists of pitch loss and expressiveness loss as shown in Eq. \ref{eq:condloss}, $\lambda_1$ and $\lambda_2$ are weight coefficients of the two losses. The pitch loss is the $L2$ distance between the predicted pitch $p'$ and the pitch label $p$, while the expressiveness loss is retrieved from the professional and modified amateur expressiveness representation.
\begin{equation}
    \mathcal L_{con} =  \lambda_1\parallel p'-p \parallel_{2} +\lambda_2\parallel W_{a}' -W_{p}\parallel_{2}
\label{eq:condloss}
\end{equation}

\subsection{Denoiser}
\begin{figure}[!ht]
    \centering
    \includegraphics[width=0.48\textwidth]{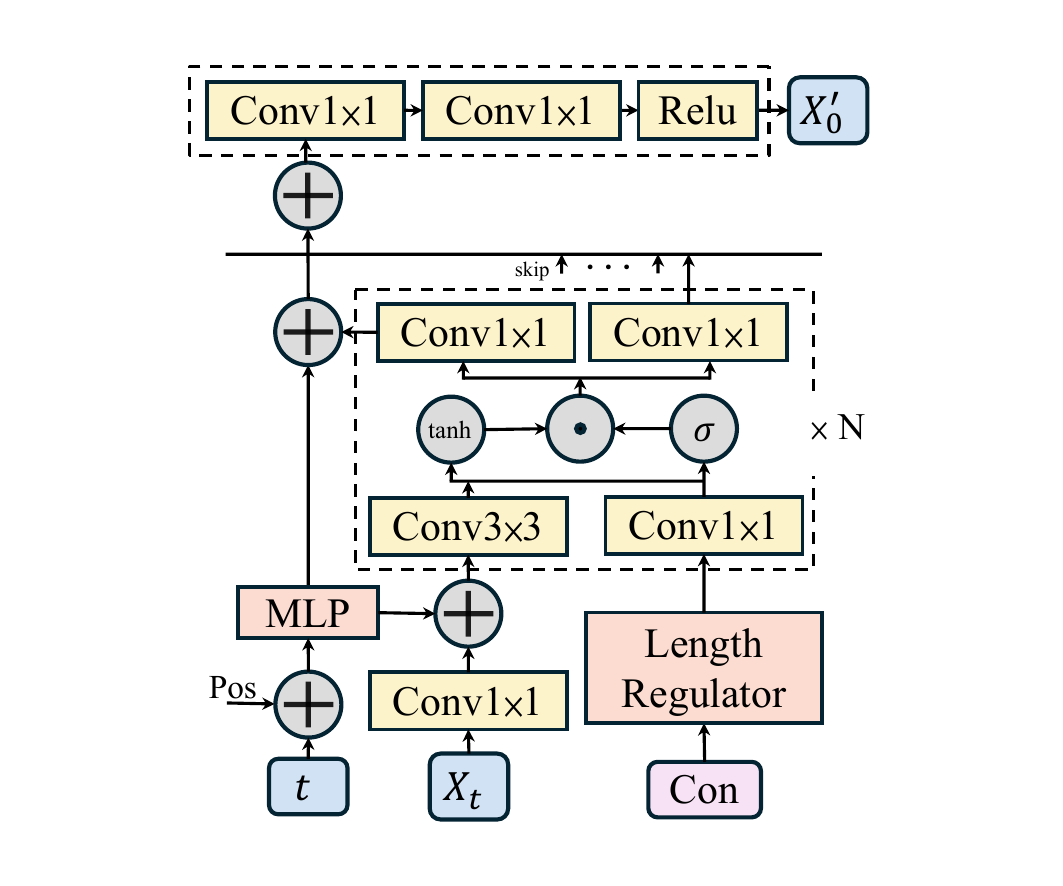}
    \caption{Structure of the spectrogram denoiser.}
    \label{FigDenoiser}
\end{figure}

Following the previous work \cite{huang2022prodiff}, we adopt a non-causal WaveNet \cite{DBLP:wavenet} architecture as our spectrogram denoiser as shown in Fig. \ref{FigDenoiser}. The denoiser comprises a $1 \times 1$ convolution layer and $\mathit{N}$ convolution blocks with residual connections to project the input hidden sequence with $256$ channels. For any step $\mathit{t}$, we use a cosine schedule with $\beta _{t}=\cos (0.5\pi t)$. The condition first passes through the length regulator to get the same dimension to $x_{t}$ through padding. Finally, the output part of the denoiser consists of a 2-layer 1D-convolutional network with ReLU activation.

\subsection{Training and Inference}
\subsubsection{Training}
The final loss term in training \textsc{ConTuner} consists of the following parts: 
\begin{itemize}
\item The reconstruction loss $\mathcal L_{\theta }^{Denoiser}$ in Eq. \ref{eq:denoiser}, which is the MSE between the generated and target Mel-spectrograms.
\item The condition loss $\mathcal L_{con}$, which is the distance between the predicted pitch and the target pitch, as well as the modified expressiveness and the professional expressiveness, as shown in Eq. \ref{eq:condloss}.
\end{itemize}
It is worth noting that the pitch predictor, expressiveness enhancer and denoiser perform gradient back-propagation at the same time during training.

\subsubsection{Inference}
\textsc{ConTuner} iteratively predicts unperturbed $x_{0}^{p}$ and then adds back perturbation via the posterior distribution. Specifically, the denoising model $f_{\theta }(x_{t}|t,con)$ predicts $x_{0}'$ firstly, and then $x_{t-1}$ is sampled using the posterior distribution $q(x_{t-1}|x_{t},x_{0}')$ that given $x_{t}$ and the predicted $x_{0}'$. In addition to the amateur singing voice, we \textbf{only} need a professional MIDI \textit{in the inference stage}.

\section{Experiment}
\subsection{Experimental Setup}
Since SVB is a novel task, with few public unaccompanied datasets so far. NSVB \cite{nsvb} comes up with PopButFy, an SVB dataset composed of paired data (\textit{i.e.} amateur and professional vocals from the same person). However, we aim to reduce the SVB model's dependency on paired data, so we collect professionally recorded songs and songs from amateur singers to produce an SVB dataset called Professional and Amateur Singing Voice dataset \textbf{PASV}. PASV consists of about 400 Mandarin and English pop songs ($\approx$42 hours) in total. In order to get closer to the scene of SVB, the amateur songs in PASV contain out-of-tune samples and extremely amateur samples. For each amateur sample, there is a one-to-one correspondence with the original professional song. For the recorded song, we use Spleeter \cite{spleeter} to separate the singing voice from the accompaniment to extract the pure human singing voice.  

We utilize the Griffin-Lim algorithm \cite{griffinlim} as the vocoder to obtain waveform from the generated Mel-spectrogram in all our experiments. \textsc{ConTuner} is trained on a 12G NVIDIA 3080Ti GPU with $400k$ steps. The warm-up learning rate is set to $10^{-4}$ and an Adam optimizer \cite{kingma2015adam} with $\beta_{1}=0.9,\beta_{2}=0.99,\epsilon=10^{-9}$ is built for training. Audio samples are available at \url{https://largeaudiomodel.com/contuner/}.


\subsection{Metrics and Measurement}
We use subjective metric mean opinion score (\textbf{MOS}, and comparison MOS, \textbf{CMOS} for ablation study) and the objective metric Mel-ceptral distortion (\textbf{MCD}) to evaluate the performance of our proposed model on the test set. In addition, following NSVB \cite{nsvb}, we also leverage pitch alignment accuracy (\textbf{PAA}) as an objective metric to measure pitch correction. For the beautified singing voices, we analyzed the MOS from two aspects: audio quality (such as naturalness, singing voice quality, \textit{etc.}, denoted as \textbf{MOS-Q}) and expressiveness (such as singing skills, emotion, rhythm, \textit{etc.}, denoted as \textbf{MOS-E}).

\subsection{Experimental Results}
\subsubsection{Pitch Correction}
\begin{figure}[!ht]
    \centering
    \includegraphics[width=0.49\textwidth]{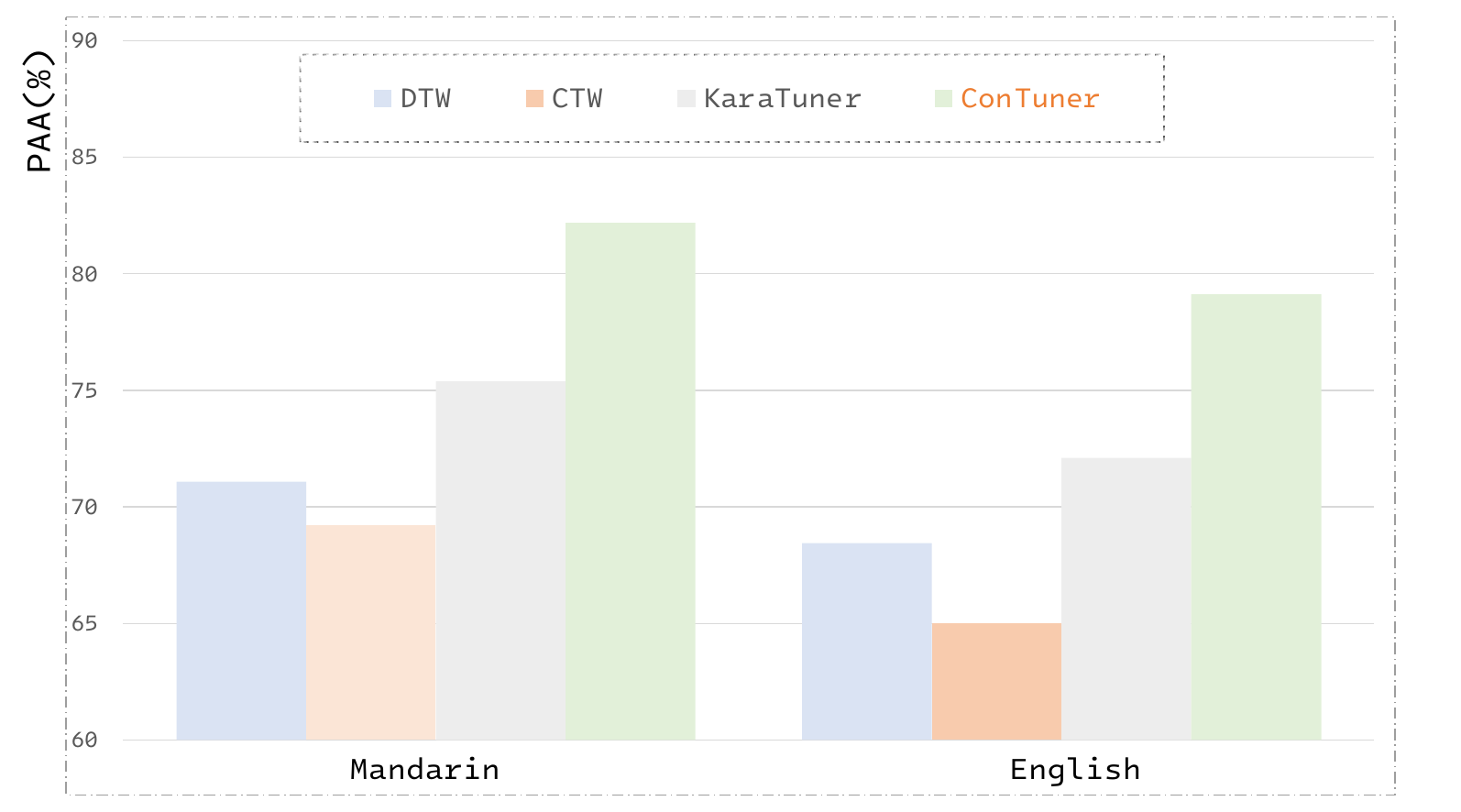}
    \caption{The pitch alignment accuracy of different algorithms on Mandarin and English songs.}
    \label{FigPAA}
\end{figure}

Fig. \ref{FigPAA} shows the results of the comparison between our proposed method and other methods on PAA. Dynamic Time Warping (DTW) \cite{DTW} and Canonical Time Warping (CTW) \cite{zhou2009canonical} are two classic algorithms for pitch correction \cite{luo2018singing}, while KaraTuner \cite{zhuang2021karatuner} is a Transformer-based method that performs pitch correction. \textsc{ConTuner}(P) represents the pitch predictor in our proposed model. Results show that PP in \textsc{ConTuner} outperforms other time-aligned methods. This is mainly because previous time-warping algorithms only focus on the forced alignment in time but ignore the direction of the pitch curve.

\subsubsection{Quality and Expressiveness}
In order to analyze the beautification capability of our proposed model, we further compare the quality MOS-Q, expressive quality MOS-E and objective metric MCD with other baseline methods. The following models and ground-truths are compared,
\begin{itemize}
\item Ground-truth Mel includes amateur (GT MelA) and professional (GT MelP) versions. We first convert ground-truth audio into Mel-spectrogram, and then convert the Mel-spectrogram back to waveform via the vocoder to eliminate the effect of vocoder on evaluation.
\item KaraTuner\cite{zhuang2021karatuner}, a neural pitch correlation model.
\item \textsc{ConTuner}, the propoesd SVB model.
\end{itemize}
Table \ref{tab:mosmcd} shows the results of \textsc{ConTuner} and other models on both Mandarin and English singing voice data. Results show that 
\begin{itemize}
\item \textsc{ConTuner} significantly achieves promising results in both MOS-Q and MOS-E, with audio quality degradation with $0.12$, as well as MOS-E being more than those for ground-truth amateur recordings by $1.08$ and $1.00$ on Mandarin and English data respectively. This result proves the strong performance of the \textsc{ConTuner} in singing voice beautification. 
\item As for MOS-E, \textsc{ConTuner} is less than those for ground-truth professional recordings by only $0.15$ and $0.21$ in Mandarin and English singing voice data respectively, which proves that the \textsc{ConTuner} has strong language generalization performance.
\end{itemize}
Comparing to other methods, \textsc{ConTuner} achieves satisfactory results. It shows that under the control of the condition, the Mel-spectrogram will be optimized with a more accurate pitch and better expressiveness.

\begin{table}[t]
    \centering
    \caption{ Comparison of different methods \\
    (with 95$\%$ confidence intervals).}
    \begin{tabular}{p{0.5cm}<{\centering}cccc}
        \toprule
        \textbf{Lang.} & \textbf{Method} & \textbf{MOS-Q $\uparrow$} & \textbf{MOS-E $\uparrow$} & \textbf{MCD $\downarrow$}
        \cr
        \midrule
        \multirow{4}{*}{\rotatebox{90}{Mandarin}} &
        GT MelA & 4.31$\pm$0.21 & 3.16$\pm$0.28 & -
        \cr
        & GT MelP& 4.42$\pm$0.18 & 4.39$\pm$0.14 & -
        \cr
        & KaraTuner \cite{zhuang2021karatuner} & 4.15$\pm$0.15 & 4.02$\pm$0.13 & 7.21$\pm$0.09
        \cr
        & \textbf{\textsc{ConTuner}} & 4.21$\pm$0.13 & 4.24$\pm$0.12 & 6.97$\pm$0.14
        \cr
        \midrule
        \multirow{4}{*}{\rotatebox{90}{English}} &
        GT MelA & 4.18$\pm$0.25 & 3.03$\pm$0.13 & -
        \cr
        & GT MelP& 4.24$\pm$0.11 & 4.25$\pm$0.11 & -
        \cr
        & KaraTuner \cite{zhuang2021karatuner}& 4.01$\pm$0.07 & 3.86$\pm$0.13 & 8.81$\pm$0.11
        \cr
        & \textbf{\textsc{ConTuner}} & 4.06$\pm$0.15 & 4.03$\pm$0.12 & 7.20$\pm$0.10
        \cr
        \bottomrule
    \end{tabular}
    \label{tab:mosmcd}
\end{table}

\subsection{Ablation Study}
The number of sampling steps is the key parameter of the diffusion model. As the number of sampling steps increases, the quality of the audio generated will also improve but results in more time-consuming. Therefore, how to balance sampling steps and audio quality is an important problem. In order to explore the influence of different sampling steps on the performance of \textsc{ConTuner}, an ablation experiment is conducted. We also compare the gradient-based diffusion and our generator-based diffusion. The comparison of the effect of sampling steps is shown in Table \ref{tab:ablation1}. 

\begin{table}[htbp]
    \centering
    \caption{Comparison of the effect of sampling steps on \textbf{MOS-Q $\uparrow$} of gradient- and generator-based diffusion model \\
    (with 95\% confidence interval).}
    \begin{tabular}{ccc}
        \toprule
        \textbf{Sampling Steps} & \textbf{Gradient-Based} & \textbf{Generator-Based}
        \cr
        \midrule
        200  & 4.23$\pm$0.11 & 4.24$\pm$0.09
        \cr
        100  & 4.11$\pm$0.06 & 4.22$\pm$0.07
        \cr
        50  & 4.02$\pm$0.10 & 4.22$\pm$0.06
        \cr
        10  & 3.65$\pm$0.08 & 4.21$\pm$0.13
        \cr
        \midrule
        GT MelA & 4.31$\pm$0.21 & 4.31$\pm$0.21
        \cr
        GT MelP & 4.42$\pm$0.18 & 4.42$\pm$0.18
        \cr
        \bottomrule
    \end{tabular}
    \label{tab:ablation1}
\end{table}

As Table \ref{tab:ablation1} shows, with several sample steps and a large distribution of noise schedule, gradient-based or generator-based diffusion models could produce high-fidelity speech samples with similar results. When the amount of sampling steps gradually decreases to $10$, the audio quality of generator-based diffusion does not decrease significantly. This indicates that \textsc{ConTuner} can weaken the inherent trade-off problem of the diffusion model to a certain extent. We hold the view that the generator-based diffusion model is free from estimating the gradient for data density, which only needs to predict unperturbed $x_{0}$ and then add back perturbation using the posterior distribution. So our generator-based diffusion model can achieve a nice beautification effect without hundreds of sample steps.

Furthermore, in order to verify the specific effects of the designed EE in the singing beautification task, we compare the performance of \textsc{ConTuner} with \textsc{ConTuner}(E) which lacks the expresseive enhancer. Comparison mean opinion score (CMOS-E and CMOS-Q) is obtained for the comparison. It can be seen from Table \ref{tab:ablation2} that the EE mainly improves MOS-E and has no significant impact on audio quality, which indicates that the EE can improve the expressiveness of singing voices, achieving the purpose of singing voice beautifying.

\begin{table}[htb]
    \centering
    \caption{Comparison of \textsc{ConTuner}(E) and \textsc{ConTuner}.}
    \begin{tabular}{cccc}
        \toprule
        \textbf{Language} & \textbf{Method} & \textbf{CMOS-Q} & \textbf{CMOS-E}
        \cr
        \midrule
        \multirow{2}{*}{Mandarin} &
        \textsc{ConTuner}(E) & -0.003 & -0.140
        \cr
        & \textsc{ConTuner} & 0.000 & 0.000
        \cr
        \midrule
        \multirow{2}{*}{English} &
        \textsc{ConTuner}(E) & -0.000 & -0.230
        \cr
        & \textsc{ConTuner} & 0.000 & 0.000
        \cr
        \bottomrule
    \end{tabular}
    \label{tab:ablation2}
\end{table}

\subsection{Discussion}
Sing voice beautifying is currently in its early stages. In this section, We analyze some limitations in our work as well as something that can be done in the future. We hope that there will be more works to promote the development of SVB.

In this paper, we only consider the beautification of singing voices in a single scene. In the future, new beautification scenarios will bring new problems. We decouple expressiveness features via a pre-trained encoder for expressiveness enhancement, this introduces some limitations to \textsc{ConTuner}. On the other hand, The establishment of the control of the condition makes the consideration of more factors like singing skills and emotions for SVB possible. This could be useful for the quest for SVB in the future works.

\section{Conclusion}
In this paper, we propose \textsc{ConTuner}, a fast diffusion-based model for high-fidelity singing voice beautifying with pitch and expressiveness conditions. The proposed model does not need paired data for inference, it can establish the control of the condition in the process of spectrogram generation. With fewer sampling steps, our model achieves a beneficial effect on singing voice beautification. Experimental results show that \textsc{ConTuner} gets satisfactory performance on PAA and outperforms the baseline method with higher quality and better expressiveness in both Mandarin and English singing voice samples.

\section{Acknowledgement}
Supported by the Key Research and Development Program of Guangdong Province (grant No. 2021B0101400003) and Corresponding author is Xulong Zhang (zhangxulong@ieee.org).

\bibliographystyle{IEEEtran.bst}
\bibliography{refs.bib}

\end{document}